\documentclass[12pt]{emulateapj}
\usepackage[]{epsfig,graphics,amssymb,natbib}

\newcommand{\beq}{\begin{equation}}
\newcommand{\eeq}{\end{equation}}
\newcommand{\beqa}{\begin{eqnarray}}
\newcommand{\eeqa}{\end{eqnarray}}
\def\imag{\dot \imath}
\def\ha{\frac{1}{2}}

\usepackage[breaklinks, colorlinks, citecolor=blue, colorlinks=true,linkcolor=blue]{hyperref} 
\usepackage{pdfcolmk} 

\begin{document}

\title{Cosmological Spectral Deconvolution}

\author{Roland de Putter}
\affiliation{NASA Jet Propulsion Laboratory, California Institute of Technology, 4800 Oak Grove Drive, MS 169-237, Pasadena, CA, 91109, U.S.A.}
\affiliation{California Institute of Technology, MC 249-17, Pasadena, California, 91125 U.S.A.}
\author{Gilbert P. Holder}
\affiliation{Department of Physics, McGill University, 3600 Rue University, Montreal, Quebec H3A 2T8, Canada}
\author{Tzu-Ching Chang}
\affiliation{IAA, Academia Sinica, P.O. Box 23-141, Taipei 10617, Taiwan}
\author{Olivier Dor\'{e}}
\affiliation{NASA Jet Propulsion Laboratory, California Institute of Technology, 4800 Oak Grove Drive, MS 169-215, Pasadena, CA, 91109, U.S.A.}
\affiliation{California Institute of Technology, MC 249-17, Pasadena, California, 91125 U.S.A.}

\begin{abstract}
One of the main goals of modern observational cosmology is to map
the large scale structure of the Universe.
A potentially powerful approach for doing this would be to exploit
three-dimensional spectral maps, i.e.~the specific intensity of extragalactic light as a function
of wavelength and direction on the sky,
to measure spatial variations in the total extragalactic light emission
and use these
as a tracer of the clustering of matter.
A main challenge is that the observed intensity as a function of wavelength
is a convolution of the source luminosity density with the rest-frame spectral energy distribution.
In this paper, we introduce the
method of spectral deconvolution as a way to
invert this convolution and extract the clustering information.
We show how one can use observations of
the mean and angular fluctuations of extragalactic light as a
function of wavelength, assuming statistical isotropy, to reconstruct jointly the rest-frame
spectral energy distribution of the
sources and the source spatial density fluctuations.
This method is more general than the well known line mapping technique as it does not rely on
spectral lines in the emitted spectra.
After introducing the general formalism, we discuss its implementation and
limitations. This formal paper sets the stage for future more practical studies.

\end{abstract}

\section{Introduction}

As spectral and imaging capabilities evolve, it is becoming increasingly
common in astronomy to think of
large three-dimensional data cubes, the specific intensity
distribution of sources along many lines of sight that
span a range of wavelengths. For studies of internal dynamics of galaxies or
molecular clouds, the frequency dimension generally corresponds to elements
at different line-of-sight velocities, while for cosmological studies
the frequency dimension (ignoring peculiar velocities) corresponds to
elements at different cosmological redshifts, and therefore at different
distances.

Cosmological large scale structure surveys routinely use
the spatial distribution of extragalactic light emission or absorption as a tracer
of the matter density to constrain cosmological models.
Such surveys
typically exploit a subset of the three-dimensional data cube
of the aggregate intensity of extragalactic light as a function of
direction on the sky and of wavelength\footnote{The caveat to this description is that it ignores polarization and
time-domain information (and cosmological information beyond the electromagnetic spectrum).},
for instance by working with a catalog of positions of a set of bright objects.
In this article, we instead consider the scenario where a
full
spectral map
is available, such as could be obtained\footnote{In realistic applications, this map would of course still be limited by
the sky coverage, wavelength range, and angular and spectral resolutions
of the survey(s) providing the data.} from
integral field spectroscopy (e.g.~HETDEX\footnote{http://hetdex.org} \citealt{hilletal08}) or narrow band imaging
surveys
(e.g.~J-PAS\footnote{http://j-pas.org/} \citealt{benitez09,molesetal10}, PAU\footnote{http://www.pausurvey.org/}
\citealt{benitez09}, or the Alhambra Survey \citealt{molesetal08}).
In other words, the data are treated at the level of a continuous three-dimensional map
as opposed to a discrete catalog.

The spectral intensity as a function of wavelength along a given line of sight in such a map
is in general a superposition of the contributions of multiple sources (and absorbers).
The goal is then to reconstruct the distribution of these sources as a function of redshift.
The difficulty here is that, generically, sources emit over a wide range of wavelengths,
so that the mapping from redshift to wavelength is degenerate.
One common strategy to evade this degeneracy is to focus on objects
that are bright enough that they stand out from the background and dominate the signal
along a given line-of-sight (such as bright galaxies or quasars),
i.e.~the ``catalog-based'' approach mentioned above.
Depending on how well the redshift direction is sampled,
this approach has led to measurements of 2-dimensional (angular projected) clustering
(e.g.~\citealt{hauserpeebles73,efstmoody01,scrantonetal02,tegmarketal02,frithetal05,coorayetal10,donosoetal13}),
``2 + 1-dimensional''
clustering using photometric redshifts
(e.g.~\citealt{padetal07,rossetal11,hoetal12,hoetal13}) and of full 3-dimensional clustering using spectroscopic redshifts
(e.g.~\citealt{FKP,coleetal05,eisetal05,beutleretal11,parkinsonetal12,delatorreetal13,Anderson:2013zyy}).
An alternative approach to reconstructing the distribution of sources is line mapping,
where the signal from well-known spectral lines, e.g.~the HI 21-cm transition,
allows a direct
mapping between wavelength and redshift (e.g.~\citealt{changetal08,loebwyithe08,visloeb10,Chang:2010jp,pritloeb12,Switzer:2013ewa}).

Here, we study the reconstruction of the source luminosity density in the more general case
where we do not restrict ourselves to bright sources nor rely on spectral lines only.
Instead, we consider reconstruction of the total luminosity density of all sources contributing
to the extragalactic signal, optimally exploiting both the spectral and angular variations in the spectral map.
While for simplicity we will assume the signal is described in terms of a single effective rest-frame
spectral energy distribution (SED) shape, this method does not rely on prior knowledge of this SED form,
but instead measures it from the
data itself.

Formally, our work stems from the following observation. Generally, the
specific intensity, $I$, at a given observed wavelength, 
$\lambda$, can be written as the weighted sum along the line of sight
of the contribution of sources of spectrum, $s$, at a a
variety of cosmological redshifts:  
\begin{equation}
\label{eq:int}
I(\ln \lambda) = \int d z \, w(z) \, s(\ln \lambda,z),
\end{equation}
where $w(z)$ is a weight function proportional to the density of
emitters. If we assume that the spectral energy distribution is
non-evolving with redshift and write it in terms of the emitted SED
$s_{rest}$,  the specific intensity can be written as the convolution
\begin{equation}
\label{eq:conviso}
I(\ln \lambda) = \int d \ln (1 + z) \, \left[(1 + z) w(z) \right] \,
s_{rest}(\ln \lambda - \ln (1 + z)). 
\end{equation}
The observation that the specific intensity is a pure convolution,
where the rest-frame SED has been convolved with the redshift
distribution, bears important implications that we explore in this
work. Since both the SED and the redshift distributions are unknown
quantities of cosmological interest,  we study in particular how this
mathematical structure can be fully exploited to reconstruct these quantities. 

Throughout this work, we will develop an explicit procedure for the spectral
deconvolution of Eq.~(\ref{eq:int}) and will elaborate on
its domain of validity. We will introduce a general formalism for
single type sources and discuss the case of an emitter of known SED in Sec.~\ref{sec:known_sed}. In
Sec.~\ref{sec:unknown_sed}, we discuss the use of angular fluctuations
in the intensity map, taking advantage of the assumption of statistical
isotropy to express the observables in terms of angular (cross-)power
spectra.
In Sec.~\ref{sec:rec_sed}, we then
show how measuring fluctuations over a sufficiently large range of
angular scales and wavelengths can potentially allow us to reconstruct
both the a priori unknown SED and the underlying density fluctuations. We
conclude in  Sec.~\ref{sec:conclusion} and discuss in more detail the
range of scales at which spectral deconvolution is feasible in Appendix \ref{sec:range validity}.

\section{Observed spectra and deconvolution from a known emitted spectrum}
\label{sec:known_sed}

We describe the observed specific intensity
in terms of two quantities. On the one hand,
the normalized rest-frame SED $s_{rest}(\ln\lambda)$ quantifies the
power as a function of wavelength $\lambda$ (to simplify later notation we write it in terms of the logarithm
of wavelength $\ln\lambda$)
emitted by a source, normalized by the luminosity of the source. It could be measured in W/\AA/$L_\odot$ (i.e.~per solar luminosity).
In other words,
the observed (redshifted) differential flux per unit wavelength for a single source at redshift $z$ with luminosity $L$,
is
\begin{equation}
S_{\lambda}(\ln\lambda) = \frac{L \, s_{rest}(\ln\lambda - \ln(1+z))}{4 \pi (1 + z) D_L^2(z)} \quad {\rm (single \, \, source)},
\end{equation}
where $D_L$ is the luminosity distance to the object.
The factor $(1+z)$ in the denominator arises because this is the flux per unit wavelength, and
wavelength intervals $d\lambda$ are redshifted.
The second ingredient is the luminosity density of sources
as a function of direction $\hat{n}$ and redshift,
$\mathcal{L}(z,\hat{n})$ (in units solar luminosity $L_\odot$ per comoving volume),
which is given by an integral over the luminosity function,
\begin{equation}
\mathcal{L}(z,\hat{n}) = \int dL \, L \, \phi(L,z,\hat{n}).
\end{equation}
Here the luminosity function $\phi(L,z,\hat{n})$ gives the number of objects per comoving volume
per luminosity interval $dL$.
For the study of cosmological large scale structure,
our main interest is in the spatial variations of $\mathcal{L}(z,\hat{n})$,
as they are expected to be a biased tracer of the underlying dark matter distribution,
thus providing information on the bias of sources and the shape of the matter power spectrum.

Putting these two ingredients together,
the specific intensity $I_\lambda$ (measured in, say, W m$^{-2}$ \AA$^{-1}$ sr$^{-1}$)
can be written as the superposition of the emission along
the line of sight at a range of cosmological redshifts,
\begin{equation}
\label{eq:conviso}
I_{\lambda}(\ln \lambda,\hat{n}) = \int d \ln (1 + z) \, {\mathcal{L}(z,\hat{n}) \over 4 \pi (1 + z)^2 H(z) } \,  s_{rest}(\ln \lambda - \ln (1 + z)).
\end{equation}
The above expression will be the basis for the
work presented in this article. 

The main assumption that goes into Eq.~(\ref{eq:conviso}) is
that the emitted SED
is independent of position (i.e.~of both $z$ and $\hat{n}$), up to a normalization, so that
angular variations are due to anisotropy in the source luminosity density
and the wavelength dependence of the specific intensity is a simple convolution.
This assumption would hold if all sources had the same SED {\it shape},
independent of location.
Alternatively, the SED shape $s_{rest}$ can be interpreted as a mean SED
averaged over different objects and luminosities. Our description would then be an appropriate one
if the relative contributions to $s_{rest}$ from different individual SED's
were independent of position. The level to which Eq.~(\ref{eq:conviso}) is a realistic description
depends on the wavelength range considered. While nature will typically be more complicated than this,
it serves as a good starting point, which can later be refined.

As noted above, a key property of the specific intensity is that Eq.~(\ref{eq:conviso})
can be written as a
convolution of the redshift distribution of emitters with the rest
frame SED.
Let us first simplify the expressions by defining
\begin{equation}
g(N,\hat{n}) \equiv \frac{\mathcal{L}(z,\hat{n})}{4 \pi (1 + z)^2 H(z)},
\end{equation}
where we have defined $N \equiv \ln(1+z)$, the number
of e-foldings of expansion.
The function $g$ will be used extensively in the remainder of this article and
can be interpreted as a rescaled luminosity density of sources (for simplicity, we will
often refer to it as luminosity density).
Eq.~(\ref{eq:conviso}) thus
reads
\begin{equation}
\label{eq:conviso g}
I_{\lambda}(\ln \lambda,\hat{n}) = \int dN \, g(N,\hat{n}) \,  s_{rest}(\ln \lambda - N).
\end{equation}
If the rest-frame SED is known, then one can simply perform the
deconvolution by working in Fourier space.
Indeed, neglecting the position dependence for now,
Fourier transforming with respect to $\ln \lambda$, and denoting the
conjugate variable as $r$ (an indicator of spectral  resolution) and
using $\tilde{X}$ to indicate Fourier transformed quantities,  we find
for the source luminosity density,
\begin{equation}
\tilde{g}(r) = { \tilde{I}_{\lambda}(r) \over \tilde{s}_{rest}(r) }.
\end{equation}
This is a cosmological analog of the
Fourier quotient method of stellar kinematics (\citealt{Simkin1974,Sargent1977}).  

From this form, we see that it is beneficial for the reconstruction
of the luminosity density to have a rest-frame SED with broad support in the Fourier domain. For
example, a delta function in frequency can be seen to be useful, corresponding
to the well-known case of line emission (e.g., 21cm, CO or CII). However, it is not
required that the SED be a single line; a uniform SED in frequency leads
to an inability to reconstruct luminosity density  for $r \ne 0$, but any SED with 
non-trivial structure (e.g., a ``break'', as is commonly used in photometric redshift surveys) will allow the deconvolution
to be done.

\section{Exploiting angular variations when the emitted spectrum is not known}
\label{sec:unknown_sed}

If the normalized rest-frame SED function $s_{\rm rest}$ is not known, then it must be estimated
from the same data. In a single line of sight, this is not possible: it 
is impossible to differentiate between density or luminosity differences at different
redshifts and differences in intrinsic emitted SED at the same redshift, but at different wavelengths.
However, one can do this measurement at multiple locations on the sky. While
different locations on the sky will have different source densities at a given
redshift, we have assumed the normalized rest-frame SED $s_{rest}$ to not vary strongly
on sufficiently large scales (see the brief discussion in Section \ref{sec:known_sed}). We illustrate below how to use this
fact in the case of a constant rest-frame SED.

Let us thus consider the spectral density as a function of wavelength {\it and} angular position on the sky,
$I_{\lambda}(\ln \lambda, \hat{n})$.
Fourier transforming Eq.~(\ref{eq:conviso}) with respect to $\ln \lambda$ then gives,
\beq
\label{eq:intFT}
\tilde{I}_{\lambda}(r,\hat{n}) = \tilde{s}_{rest}(r) \times \tilde{g}(r, \hat{n}).
\eeq
The luminosity density function 
$g(N,\hat{n})$ is
proportional to the source density and we can thus describe the statistics of
the spatial fluctuations in this function in terms of the statistics of the clustering
of the source galaxies.
However, we cannot assume that the fluctuations
in $g(N,\hat{n})$ obey homogeneity because the background value $g(N) \equiv \langle g(N,\hat{n})\rangle$
has a redshift dependence,
and because, as we look towards higher $z$, the clustering amplitude of the sources may vary.
In general, the only applicable symmetry is isotropy.
The underlying reason is that we are studying fields defined on our past light cone, that are a function of angle on the sky
and of redshift. Their statistics are invariant under rotations, but not under shifts in the radial direction.
It makes sense then to analyze the
perturbations in the observed intensity field in terms of spherical harmonics, which we do below.

Let us write for the (rescaled) luminosity density,
\beq
g(\hat{n},N) = g(N) \left[ 1 + b(N) \delta_m(\vec{x}, N) + \epsilon \right].
\eeq
Here, ${g}(N)$ is the background value,
and $\delta_m(\vec{x}, N)$ is the homogeneous (at a fixed cosmic time)
three-dimensional {\it matter} overdensity field in comoving coordinates and the second 
argument denotes the time dependence of the field.
The position $\vec{x}$ is implicitly chosen to be on the past light cone,
i.e.~$\vec{x} = D(N) \, \hat{n}$ for a spatially flat universe, where $D(N)$
is the comoving distance to redshift $z$ with $N=\ln(1+z)$.
The quantity $b(N)$ is the luminosity weighted bias of the source density relative to the
matter density\footnote{$b(z) = \int dL \, L \, \phi(L,z) b_s(L,z)/\left( \int dL \, L \, \phi(L,z)  \right)$,
where $\phi$ is the luminosity function of sources and $b_s(L,z)$ is the bias relative to matter
of the density of sources with luminosity $L$ at redshift $z$. For simplicity, we consider large,
linear scales so that a scale-independent bias is appropriate.}.
This bias may evolve both due to variation with redshift of the source luminosity
function, and due to
evolution of the galaxy bias at fixed luminosity.
Finally, $\epsilon$ is a stochastic shot noise contribution due to the finite number of sources
contributing to the signal. We will ignore its contribution from here on.

Restricting the analysis for simplicity to large scales which are in the linear regime,
we define the statistics of the matter overdensity field in Fourier space by
\beq
\langle \delta_m(\vec{k}, N) \delta_m(\vec{k}', N') \rangle = (2 \pi)^3 \, \delta^D(\vec{k} + \vec{k}') P_0(k) \, T(N) \, T(N'),
\eeq
where $P_0(k)$ is the matter power spectrum at $z = 0$ and $T(N)$ is
the linear growth rate of matter perturbations relative to $z = 0$ (assumed to be
scale-independent).

We can now write variations in the Fourier-space intensity field in Eq.~(\ref{eq:intFT}) as
\beq
\delta \tilde{I}_{\lambda}(r, \hat{n}) = \tilde{s}_{rest}(r) \, \int dN \, e^{\imag N r} \, g(N) \, b(N) \, \delta_m(\hat{n}, N).
\eeq
For a given $r$, this is thus a line-of-sight integral of the overdensity field, multiplied by some kernel that only depends on
$N$. This is analogous to other cosmological observables that are functions of $\hat{n}$, like the cosmic shear/convergence field,
or the overdensity of galaxies in a redshift bin. We can thus expand the $\hat{n}$-dependence in terms of spherical harmonics,
and apply the usual machinery to derive an expression for the angular
power spectrum. This leads to
\beqa
\label{eq:cl}
\lefteqn{C_\ell(r, r') = }\nonumber \\
&& \tilde{s}_{rest}^{}(r) \, \tilde{s}^*_{rest}(r') \,\frac{2}{\pi} \int dk \, k^2 \, P_0(k) \, W_\ell(k, r) \, W_\ell^*(k,r'), \nonumber \\
&& \\
&{\rm with}& \, W_\ell(k, r) = \int dN \, j_\ell[k D(N)] \, e^{\imag N r} \, 
{g}_0(N) \, b(N) \, T(N) \nonumber
\eeqa
where $j_\ell$ is the spherical Bessel function and $D(N)$ is comoving distance.
While we will neglect this term in the following, for completeness we note that, if we assume the sources
to be Poisson tracers of the underlying matter distribution, there will be a shot noise contribution
\beqa
\label{eq:cl sn}
\lefteqn{C_\ell ^{sn} (r, r') =  \tilde{s}_{rest}^{}(r) \,  \tilde{s}^*_{rest}(r')\, }&&\nonumber\\
& \times & \int dN\,
{e^{\imag N (r- r')}\over D^2(N) D'(N)}\, {g}^2(N) \left(\frac{\int dL \, L^2 \, \phi(L,N)}{\left( \int dL \, L \, \phi(L,N)\right)^2} \right), \nonumber\\
\eeqa
where $\phi$ is the luminosity function.

Returning to the contribution due to large scale clustering,
in the limit that the scale $\sim r^{-1}$ of radial/wavelength fluctuations is much larger than the transverse
fluctuation scale $\sim \ell^{-1}$, we can apply the Limber approximation, leading to
\beqa
\label{eq:cl limber}
\lefteqn{C_\ell(r, r') =  \tilde{s}_{rest}^{}(r) \,  \tilde{s}^*_{rest}(r')\, }&&\nonumber\\
& \times & \int dN\, P_0\left(\frac{\ell+\frac{1}{2}}{D(N)} \right)\,
{e^{\imag N (r- r')}\over D^2(N) D'(N)}\, {g}^2(N) \,b^2(N) \, T^2(N) \nonumber\\
\eeqa
where $D' \equiv dD/dN$.
We will discuss the range of validity of the Limber approximation in more detail in
the Appendix.

We have ignored redshift space distortions in the above discussion. While these
are in general non-negligible, they are small in the same limit where the Limber approximation
is valid. The reason for this is simply that in the Limber approximation only transverse
density modes contribute to the observed angular power spectra and that peculiar velocities due to transverse modes
do not have a line-of-sight component and thus do not cause
redshift space distortions.
Since the main result of this paper, i.e.~the procedure (discussed in the next section)
for simultaneously extracting the distribution
of sources and the rest-frame SED from the data itself, is only applied on scales where the Limber approximation
is valid, it is a justified approximation to neglect redshift space distortions.

By angle-averaging, we can also estimate the mean intensity
\beq
\label{eq:Ibar}
\tilde{I}_{\lambda}(r) = \tilde{s}_{rest}(r) \times \tilde{g}(r).
\eeq
We thus fundamentally have two observables: the mean,
Eq.~(\ref{eq:Ibar}), and the angular cross- and power-spectra,
Eq.~(\ref{eq:cl}).
In the next section, we will discuss how to extract the clustering of sources
from these data, without assuming prior knowledge of either the rest-frame SED
$\tilde{s}_{\rm rest}(\ln\lambda)$ or the mean luminosity density $g(N)$.

\section{Jointly reconstructing a constant rest-frame SED and the clustering
  of sources}
\label{sec:rec_sed}

Using the tools developed in the previous section, we can now study to what extent we can extract
interesting physical quantities from the observables $C_\ell(r,r')$ and $\tilde{I}_{\lambda}(r)$.
We would especially like to obtain a measurement of the clustering of sources,
i.e.~the information contained in the matter power spectrum
$P_0(k)$, and the biased transfer function $b(N)T(N)$. It turns out it
is possible to indeed isolate this clustering information, while
simultaneously estimating the normalized rest-frame SED $s_{\rm
  rest}(\ln\lambda)$ and the mean luminosity density $g(N)$. We explain this below.

Let us consider first the diagonal angular spectra\footnote{We expect a large fraction of the information
to reside in the $r= r'$ configurations as the terms other than the $e^{\imag N(r- r')}$ in
the integrand on the right hand side of Eq.~(\ref{eq:cl limber}) are slowly varying, so that the signal quickly
declines for $r\neq r'$. },
$r' = r$.
The observed spectrum is then a separable function of $r$ and $\ell$,
\beq
\label{eq:sep}
C_\ell(r) \equiv C_\ell(r,r) = A(\ell) \, |\tilde{s}_{rest}(r)|^2,
\eeq
with
\beq
\label{eq:A(l)}
A(\ell) \equiv \int dN \, P_0\left(\frac{\ell+\frac{1}{2}}{D(N)}
\right) \, {g^2(N) \, b^2(N) \, T^2(N)\over D^2(N) D'(N)}.
\eeq
This separability property will prove to be crucial for the argument below.

It is straightforward to obtain clustering information that is independent of the rest-frame
SED $s_{\rm rest}(\ln\lambda)$ from the diagonal spectra $C_\ell(r)$.
For instance, we can take the ratio with the squared mean intensity
defined in Eq.~(\ref{eq:Ibar}),
\beq
\label{eq:ratio}
\frac{C_\ell(r)}{|\tilde{I}_{\lambda}(r)|^2} = \frac{A(\ell)}{|\tilde{g}(r)|^2}.
\eeq
This quantity contains an integral over the clustering of sources (inside $A(\ell)$)
and otherwise only depends on the unknown function $g(N)$. Alternatively, $A(\ell)$ could have been
directly estimated using the separability of Eq.~(\ref{eq:sep}). In either case, the next challenge
is to estimate $g(N)$ from the data. This can be done in several ways and we outline a step-by-step procedure below.

\begin{itemize}
\item
First, we estimate $\tilde{s}_{\rm rest}(r)$, up to an $r$-independent normalization.
We fix this normalization by specifying $\tilde{s}_{\rm rest}(r)$ at some $r = r_0$,
i.e.~$\tilde{s}_0 \equiv \tilde{s}_{\rm rest}(r_0)$.
It turns out that this normalization does not matter for the reconstruction of the clustering
information so it is not a problem if it remains unknown. The first step is to use the separability of $C_\ell(r)$ to estimate the
norm of $\tilde{s}_{\rm rest}(r)$ for all $r$ (we will discuss the range of validity
of this procedure in more detail in Appendix \ref{sec:range validity}). Explicitly,
\beq
\label{eq:ests2}
|\tilde{s}_{\rm rest}(r)|^2 = |\tilde{s}_0|^2 \, \sum_\ell W^{(\tilde{s})}_\ell(r) \frac{C_\ell(r)}{C_\ell(r_0)}.
\eeq
Here, for each $\ell$, the quantity  $|\tilde{s}_0|^2 \, C_\ell(r)/C_\ell(r_0)$
is an estimator of $|\tilde{s}_{\rm rest}(r)|^2$ so a general estimator is written as a sum
over all multipoles with a set of weights $W^{(\tilde{s})}_\ell(r)$. In principle, these weights can be adjusted
to optimize the signal-to-noise ratio of the estimator, but we leave this question for future work.
\item
The above calculation only gives us the norm of $\tilde{s}_{\rm rest}(r)$ as a function of $r$,
but not the phase. To extract the phase information, the angular {\it cross} $r' \neq r$ power spectra
can be used. For example, we can take ratios of the following kind to estimate
\beq
\label{eq:phase}
\frac{\tilde{s}_{\rm rest}(r')}{\tilde{s}_{\rm rest}(r)} = \frac{C_\ell(r',\ha (r+r'))}{C_\ell(r,\ha (r+r'))}.
\eeq
Using this relation for various pairs of $r$ and $r'$ gives the evolution of the phase of $\tilde{s}_{\rm rest}$
(relative to the assumed phase $\tilde{s}_0 = \tilde{s}_{\rm rest}(r_0)$). In principle, these ratios also
give information on the norm, but we expect that the diagonal angular power spectra discussed in the previous step
carry more information.
\item
Now that we know $\tilde{s}_{\rm rest}(r)$, we can use the mean intensity $\tilde{I}_{\lambda}(r) = \tilde{s}_{\rm rest}(r)
\, \tilde{g}(r)$ to estimate $\tilde{g}(r)$, i.e.
\beq
\tilde{g}(r) = \frac{\tilde{I}_{\lambda}(r)}{\tilde{s}_{\rm rest}(r)}.
\eeq
Since we had to assume $\tilde{s}_0$ for our estimate of $\tilde{s}_{\rm rest}(r)$,
we have really estimated $\tilde{g}(r) \times \tilde{s}_0$.
\item
We can now insert the estimate of $\tilde{g}(r)$ (and therefore $g(N)$) into, e.g.,
Eq.~(\ref{eq:ratio}). We see that the normalization $\tilde{s}_0$ conveniently drops out
and we have now isolated the effect of $P_0(k)$ and $b(N) \, T(N)$.
\end{itemize}

Of course, we still only measure the projected power spectrum (see Eq.~(\ref{eq:A(l)})), and
uncertainty in the background
cosmology, which determines the projection through $D(N)$ and $D'(N)$, still
needs to be taken into account.

Moreover, the above relies crucially on the Limber approximation, as in general (see Eq.~(\ref{eq:cl})),
the $\ell$ and $r$ dependences do not factorize.
We therefore discuss in the Appendix for what range of multipoles $\ell$
and wavelength Fourier conjugate $r$ the above procedure can be applied.
The main conclusion (but see the Appendix for details)
is that the reconstruction described above is valid
for multipoles
$\ell \approx 30 - \ell_{\rm max}$, where $\ell_{\rm max}$ is the largest multipole that
can be observed given the angular resolution (and noise and non-linear cutoff) of the experiment.
For a given multipole $\ell$, the approach can be applied to
line-of-sight/wavelength Fourier modes $r \approx r_{\rm min} - {\rm min}(\ell/10,r_{\rm max})$,
where $r_{\rm min}$ is the minimum observed mode, set approximately by $1/\Delta \ln\lambda$,
with $\Delta \ln\lambda$ the wavelength range covered, and $r_{\rm max}$
is the maximum resolved mode, set by the spectral resolution $R$.
Put simply, the approach sketched in this work can be applied to mostly transverse modes, i.e.~those that vary more rapidly in
the transverse (angular) direction than in the line-of-sight/wavelength direction.

In this section, we have outlined an explicit series of steps that can be followed to obtain
information on the clustering of sources and on the rest-frame SED simultaneously.
When applied to data, it may make more sense to either model or parametrize the functions that
are to be reconstructed, i.e.~$g(N)$, $b(N) T(N)$, $s_{\rm rest}(\ln\lambda)$ and $P_0(k)$,
and to simultaneously fit the resulting parameters to
the 3D spectral mapping observables discussed above.
The explicit reconstruction method discussed in this section then shows that, because of statistical isotropy,
the degeneracy between these parameters
can be broken when the mean and angular variations in the map are used,
so that parameters describing the clustering on the one hand, and parameters describing
the mean luminosity and the rest-frame SED on the other hand, can in principle be constrained independently.

\section{Conclusions}

\label{sec:conclusion}

We have introduced a formalism for measuring the projected two-point function of the luminosity
density of extragalactic sources from spectral mapping data,
i.e.~from a three-dimensional data cube of specific intensity as a function
of wavelength and line of sight direction.
This method does not rely on the use of spectral lines
and works for an arbitrary rest-frame SED form.
In fact, the source SED does not even need to be known in advance,
but is reconstructed from the data itself.

This spectral deconvolution technique makes use of both the mean intensity as a function of wavelength
and its angular variations.
The wavelength dependence of the rest-frame SED
and the redshift dependence of the luminosity density of sources
can be disentangled
because the observed specific intensity is a convolution of these two quantities.
This allows a straightforward spectral deconvolution in Fourier space.
The spatial variations in the reconstructed luminosity density contain valuable information
on the large scale clustering of matter.

While one might expect this approach to only work if the (mean) rest-frame SED
is known a priori, we have shown in Section \ref{sec:rec_sed}
how to use the statistical isotropy of the observed signal
to simultaneously measure the rest-frame SED and the clustering of
the luminosity density when this is not the case. 
Specifically, by first Fourier transforming the
spectral map with respect to $\ln\lambda$ (with Fourier conjugate $r$),
and then considering
the angular power spectrum for mostly transverse modes (multipoles $\ell >> r$),
the term describing the clustering of extragalactic sources to a good approximation
only depends on multipole $\ell$, while the rest-frame SED term only depends on the wavelength direction $r$.
It is this separability of variables that allows for their independent reconstruction.

The purpose of this paper has been to give a rather formal presentation
of the method of spectral deconvolution,
leaving more concrete explorations of how to apply the technique to
realistic data for future work.
A particularly strong assumption we have made throughout this work
is that the extragalactic signal can be described in terms of a single effective
rest-frame SED (up to a free normalization). It will be interesting to generalize the method
to scenarios where multiple populations with distinct SED's need to be factored in.

One additional motivation for studying the information content of three-dimensional spectral maps,
beyond the fact that such data will be available from spectral mapping experiments,
is that, as discussed in the Introduction, such maps provide a unifying description
for a large range of cosmological surveys, with different probes accessing different subsets
of the data cube.
It would thus be useful to build a more general understanding of all the information
that can in principle be extracted from the full data cube. This can then be a guide towards identifying how to optimally
exploit these data in the future.

\acknowledgments{
We acknowledge the Aspen Center for Physics where part is this work
originated. The center is supported by NSF grant 1066293. TC
acknowledges support from the Simons Foundation.  Part of the research
described in this paper was carried out at the Jet Propulsion
Laboratory, California Institute of Technology, under a contract with
the National Aeronautics and Space Administration.}

\appendix

\section{Angular and wavelength validity range}
\label{sec:range validity}

Section \ref{sec:rec_sed} in principle gives us a step-by-step approach to
reconstructing the clustering and rest-frame SED from the spectral mapping data.
We here estimate for what range of multipoles $\ell$ and line of sight wave numbers
$r$ this procedure is applicable.
We assume the angle averaged intensity $\bar{\tilde{I}}(r)$ and the anisotropies
in the intensity $\delta \tilde{I}_{\lambda}(r, \hat{n})$ can be reliably measured for some range
\beq
\label{eq:Lambda range}
r\in \left[ r_{\rm min}, r_{\rm max} \right],
\eeq
\beq
\label{eq:ell range}
\ell \in \left[ \ell_{\rm min}, \ell_{\rm max} \right].
\eeq
While it goes against the spirit of this paper to precisely quantify these bounds
for a realistic experiment,
their interpretation in terms of survey properties is easily understood.
The lower bound $r_{\rm min}$ is related to the range of wavelengths
for which we can observe the spectrum, $r_{\rm min} \sim 1/\Delta \ln \lambda$
(with $\Delta \ln \lambda$ the observed spectral range in $\ln \lambda$),
while $r_{\rm max}$ is related to the spectral resolution of the instrument
$r_{\rm max} \sim 1/d\ln \lambda \sim R$ (with $d \ln \lambda$ the wavelength resolution).
For the multipole range, $\ell_{\rm min}$ is given by the sky coverage of the survey,
while $\ell_{\rm max}$ would be determined by the resolution and by the angular scale
at which the noise becomes large. Moreover, restricting the analysis to clustering
in the linear regime places a constraint on the maximum multipole that can be used.

Another important restriction on the range of phase space that can be used comes from the
Limber approximation, which was a crucial assumption in the derivations of the previous section.
The Limber approximation is valid for multipoles larger than a critical scale, which may depend
on $r$,
i.e.~$\ell > \ell_{L}(r)$.
To quantify $\ell_{L}(r)$, consider the analysis
presented in \cite{LoverdeAfshordi2008}.
Here, an explicit expression for the leading order correction
to the Limber approximation is presented and used to quantify
the range of validity of the Limber approximation.
In their notation, the integrated kernel relevant for the spectral mapping scenario
studied in the present paper can be written as
\beq
\label{eq:kernel}
f_{A/B}(\chi) = \frac{a H}{\sqrt{\chi}} \, g(\chi) \, b(\chi) \, T(\chi) \, e^{\pm \imag N(\chi) r} \equiv \bar{f}(\chi) \, e^{\pm \imag N(\chi) r}.
\eeq
To avoid confusion with the ``resolution parameter'' $r$ (the Fourier conjugate of $\ln\lambda$),
we have used the letter $\chi$ for comoving distance instead of $r$ (which is used in \citealt{LoverdeAfshordi2008}).
We have explicitly expressed ``line-of-sight'' quantities in terms of $\chi$ instead of $N=\ln(1+z)$.

In Eq.~(13) of \cite{LoverdeAfshordi2008}, the correction to the Limber approximation is expressed as an integral
over $d\ln f_A/d\ln \chi \, d\ln f_B/d\ln \chi$. The correction is small, and therefore the Limber approximation
good, if the product of this quantity with $\ell^2$ is small, which in our case means
\beqa
\ell^{-2} \, d\ln f_A/d\ln \chi \, d\ln f_B/d\ln \chi &=& \nonumber \\
\ell^{-2} \, \left( \left(d\ln \bar{f}/d\ln \chi \right)^2 + (a H \chi)^2 r^2 \right) &\ll& 1.
\eeqa
We require that both terms in the parentheses satisfy the condition individually.
The first term in the parentheses then gives the usual condition appropriate for a kernel
$\bar{f}(\chi)$, say $\ell > \ell_{L,0}$.
We expect the kernel $\bar{f}(\chi)$ relevant for spectral mapping to be relatively wide because
the luminosity density $g(\chi)$, the product of bias and transfer function $b(\chi) \, T(\chi)$,
and the geometric factors in Eq.~(\ref{eq:kernel}), should all vary slowly with distance.
Assuming therefore that the mean distance $\bar{\chi}$ of the kernel and its width $\Delta \chi$
are of the same order of magnitude, we expect $\left(d\ln \bar{f}/d\ln \chi \right)^2$
to be of order unity. To be conservative, we then use
$\ell_{L,0} \sim 30$. The second term in the parentheses gives an $r$-dependent requirement,
$\ell > \ell_{L, r} \, r$ and we estimate $\ell_{L, r} \sim 10$ (assuming the typical redshift for which the kernel
is large to be $z \sim 1$).
Summarized, we get the following constraint:
\beqa
\label{eq:limber req}
\ell > \ell_{L}(r) &\equiv & {\rm max} \{ \ell_{L,0}, \ell_{L,r} \, r \}\\
\rm{with}\quad \ell_{L,0} \sim 30 &\quad {\rm and}\quad & \ell_{L,r} \sim 10 \nonumber.
\eeqa

Equations (\ref{eq:Lambda range}), (\ref{eq:ell range}) and (\ref{eq:limber req})
list the restrictions on the range of scales for which we can use $C_\ell(r)$. To continue, we assume
\beqa
\ell_{\rm min} &<& \ell_{L,0} < \ell_{\rm max}, \\
r_{\rm min} &<& \frac{\ell_{L,r}}{\ell_{L,0}}. 
\eeqa
These are realistic assumptions and simplify the following expressions.
For a given $r$ in the range (\ref{eq:Lambda range}), we can use the
$C_{\ell}(r)$'s in the range $\ell \in \left[ \ell_L(r), \ell_{\rm max}\right]$.
Requiring that this range is non-zero gives the range of $r$ that is usable,
\beqa
\label{eq:range Lambda}
r&\in& \left[ r_{\rm min}, {\rm min}\{ \ell_{\rm max}/\ell_{L,r}, r_{\rm max} \} \right] \nonumber \\
\ell &\in& \left[ \ell_L(r), \ell_{\rm max}\right],
\eeqa
or equivalently
\beqa
\label{eq:range ell}
\ell &\in& \left[ \ell_{L,0}, \ell_{\rm max}\right] \nonumber \\
r&\in& \left[ r_{\rm min}, {\rm min}\{ \ell/\ell_{L,r}, r_{\rm max} \} \right].
\eeqa
The usable range for the angle-averaged intensity is simply given by Eq.~(\ref{eq:Lambda range}).

Given the range of scales where we can use the observables $C_\ell(r)$ and $\tilde{I}_{\lambda}(r)$,
we can now write the
estimator for $|\tilde{s}_{\rm rest}(r)|$, given in
Eq.~(\ref{eq:ests2}) in the previous section, as
\beq
\label{eq:ests2}
|\tilde{s}_{\rm rest}(r)|^2 = |\tilde{s}_0|^2 \, \sum_{\ell \in \left[\ell_L(r), \ell_{\rm max} \right]} W^{(\tilde{s})}_\ell(r) \frac{C_\ell(r)}{C_\ell(r_0)},
\eeq
where we have now explicitly specified the range of multipoles to be summed over.
For the optimal range of $\ell$ to be available, we need to choose
\beq
r_0 \in \left[ r_{\rm min}, \frac{\ell_{L,0}}{\ell_{L,r}} \right].
\eeq

Following the entire series of steps detailed in Section \ref{sec:rec_sed},
and applying the restrictions on the range of scales that can be used,
as given in Eq.~(\ref{eq:range Lambda}) or (\ref{eq:range ell}),
we conclude that the projected power spectrum of the source density,
$A(\ell)$ can be measured for all $\ell$ given in Eq.~(\ref{eq:range ell}).
The background luminosity density $\tilde{g}(r)$ and rest-frame SED $\tilde{s}_{\rm rest}(r)$
can both be reconstructed for all $r$
in Eq.~(\ref{eq:range Lambda}).

To extract the clustering information
from $A(\ell)$, we want to insert the function $g(N)$, which is obtained by
Fourier transforming the reconstructed function $\tilde{g}(r)$. The range of $r$,
for which this reconstruction is possible tells us that we can only
reconstruct variations in $g(N)$ on scales smaller than $(\Delta N)_{\rm max} \sim 1/r_{\rm min} \sim \Delta \ln\lambda$
(the wavelength range covered by the spectral mapper). Moreover, we cannot recover variations on scales
smaller than $(\Delta N)_{\rm min} \sim {\rm max}(\ell_{L,r}/\ell_{\rm max},1/r_{\rm max} \sim R^{-1})$,
which is the maximum of the wavelength resolution and $\sim 10 \times $ the angular resolution scale (in radians).
This will limit how well one can recover the power spectrum information.

\bibliography{im}

\end{document}